\documentclass[usenatbib]{emulateapj}
\usepackage{graphicx}
\usepackage[flushleft]{threeparttable}
\usepackage[usenames,dvipsnames,svgnames,table]{xcolor}
\usepackage{hyperref}
\definecolor{darkblue}{rgb}{0.0,0.0,0.3}
\hypersetup{colorlinks,breaklinks,
            linkcolor=darkblue,urlcolor=darkblue,
            anchorcolor=darkblue,citecolor=darkblue}
\usepackage{amsmath,amssymb}
\usepackage{amsmath}

\usepackage{natbib}
\makeatletter
\renewcommand{\p@subsection}{}
\renewcommand{\p@subsubsection}{}
\makeatother

\begin{document}

\def\etal{et al.\ \rm}
\def\ba{\begin{eqnarray}}
\def\ea{\end{eqnarray}}
\def\etal{et al.\ \rm}
\def\Fdw{F_{\rm dw}}
\def\Tex{T_{\rm ex}}
\def\Fdis{F_{\rm dw,dis}}
\def\Fnu{F_\nu}
\def\WD{\rm WD}

\newcommand\cmtrr[1]{{\color{red}[RR: #1]}}
\newcommand\cmtla[1]{{\color{blue}[LA: #1]}}


\title{Spin Evolution and Cometary Interpretation of the Interstellar Minor Object 1I/2017 'Oumuamua}

\author{Roman R. Rafikov\altaffilmark{1,2}}
\altaffiltext{1}{Centre for Mathematical Sciences, Department of Applied Mathematics and Theoretical Physics, University of Cambridge, Wilberforce Road, Cambridge CB3 0WA, UK; rrr@damtp.cam.ac.uk}
\altaffiltext{2}{Institute for Advanced Study, Einstein Drive, Princeton, NJ 08540}


\begin{abstract}
Observations of the first interstellar minor object 1I/2017 'Oumuamua did not reveal direct signs of outgassing that would have been natural if it had volatile-rich composition. However, a recent measurement by Micheli et al (2018) of a substantial non-gravitational acceleration affecting the orbit of this object has been interpreted as resulting from its cometary activity, which must be rather vigorous. Here we critically re-assess this interpretation by exploring the implications of measured non-gravitational acceleration for the 'Oumuamua's rotational state. We show that outgassing torques should drive rapid evolution of 'Oumuamua's spin (on a timescale of a few days), assuming torque asymmetry typical for the Solar System comets. However, given the highly elongated shape of the object, its torque asymmetry is likely higher, implying even faster evolution. This would have resulted in rapid rotational fission of 'Oumuamua during its journey through the Solar System and is clearly incompatible with the relative stability of its rotational state inferred from photometric variability. Based on these arguments, as well as the lack of direct signs of outgassing, we conclude that the classification of 'Oumuamua as a comet (invoked to explain its claimed anomalous acceleration) is questionable.
\end{abstract}


\keywords{planetary systems --- minor planets, asteroids: general --- minor planets, asteroids: individual ('Oumuamua)}


\section{Introduction.}  
\label{sect:intro}


Discovery of 1I/2017 'Oumuamua \citep{Meech} --- the first macroscopic interstellar object passing through the Solar System --- by the Pan-STARRS survey \citep{Chambers} gave rise to a number of questions and theories of its origin \citep{Trilling,Raymond,Cuk,Hansen,Jackson,Rafikov}. Physical characteristics of 'Oumuamua, namely its composition --- refractory or volatile-rich --- play a central role in this debate. 

'Oumuamua was found to not exhibit the standard, obvious signs of cometary activity \citep{Meech,Knight17,Jewitt,Ye,Fitz} --- outgassing, formation of coma and dust tails, cometary emission lines, etc. Its colors suggest that 'Oumuamua might look like a volatile-rich object that suffered long exposure to cosmic rays. Nevertheless, until recently it was mainly the formation arguments (rooted in the idea that it is easier to eject material from the outer, cold regions of the planetary systems) that still motivated the possibility of volatile-rich composition for 'Oumuamua \citep{Batygin}.

Recently, this hypothesis has received additional (albeit indirect) support. \citet{Micheli} used long term astrometric measurements of 'Oumuamua's orbit to claim the detection of a substantial non-gravitational acceleration affecting its motion. This acceleration is predominantly radial, and has been modeled reasonably accurately by \citet{Micheli} as a function of distance $r$ from the Sun (in the range $1$ au $<r< 3$ au) as
\ba  
a_{\rm ng}(r) &=& A_1g(r),~~~~~g(r)=r_1^{-2},
\label{eq:accel}\\
A_1 &\approx & 5\times 10^{-4}~\mbox{cm s}^{-2}=2.5\times 10^{-7}\mbox{au d}^{-2},\nonumber
\ea   
where $r_1=r/1$ au.

Anomalous accelerations (unaccounted for by the Solar gravity) are known to affect orbits of many Solar System comets \citep{Krolik,Szut}. They are naturally ascribed to the reactive force due to mass loss (outgassing) powered by the Solar heating. For this reason, the discovery of anomalous acceleration of 'Oumuamua led \citet{Micheli} to conclude that it must be a volatile-rich object capable of exhibiting cometary activity. At the same time, no other, more direct, manifestations of the cometary activity of 'Oumuamua  were found by either \citet{Micheli} or others. To reconcile these mutually exclusive lines of evidence \citet{Micheli} had to suggest rather exotic scenario, in which mass loss occurs mainly in the form of gas (with almost no ejection of small dust particles) with composition very different from that of the Solar System comets.

In this note we explore the implications of the claimed substantial non-gravitational acceleration of 'Oumuamua for its rotational state. The observed spin state of 'Oumuamua is far from trivial (see \S \ref{sect:rot}), however, most recent observations find its rotational period to remain close to $8$ hr over month-long interval \citep{Bolin,Bannister,Drahus,Fraser,Belton}. In this work we find that, given the magnitude of the claimed non-gravitational acceleration of 'Oumuamua, this stability of its spin period is highly unusual (\S \ref{sect:spin_Ou}). This observation represents yet another argument putting into question the interpretation of 'Oumuamua as a comet. Our calculations make extensive use of the recent results \citep{RRR} on the connection between the outgassing-driven spin evolution and non-gravitational acceleration of the Solar System comets (\S \ref{sect:spin}).


\section{Rotational state of 'Oumuamua}  
\label{sect:rot}


Here we provide a brief overview of existing results on the spin state of 'Oumuamua. Initial measurements of its photometric variability by different groups did not find a well-defined spin period $P$ for this object, suggesting values in the interval $6.8-8.7$ hr \citep{Bolin,Bannister,Drahus,Fraser}. This ambiguity is caused by the tumbling motion of 'Oumuamua --- its non-principal axis rotation, which complicates the interpretation of its photometric variability. More recently \citet{Belton} analyzed different photometric datasets spanning about a month in duration(from October 25, 2017 to November 23, 2017) and found a set of prominent frequencies in these time series. The dominant period of $(8.67\pm 0.34)$ hr was associated by \citet{Belton} with the precession of the 'Oumuamua's long axis. Other periodicities, representing nutation and rotation around the long axis depend on the interpretation of the dominant mode of the 'Oumuamua's rotation (long or short axis modes).

In this work we adopt the conventional interpretation of 'Oumuamua as being in the short axis mode of rotation \citep{Meech,Jewitt,Bolin}. In this case the high amplitude of its photometric variability is explained by 'Oumuamua's unusual, highly elongated shape. We describe it as a highly-prolate ellipsoid with semi-axes (for albedo $p=0.1$) $a\times b\times c=$230 m$\times$35 m$\times$35 m \citep{Jewitt}. For simplicity, we ignore the complications related to the excited spin state of 'Oumuamua and approximate it as rotation around its short axis with period $P=8.67$ hrs. The moment of inertia relative to this axis should be close to $I=(1/5)M(a^2+b^2)$, where $M$ is the object mass.


\section{Spin evolution due to cometary activity}  
\label{sect:spin}


Our goal is to understand the impact of comet-like outgassing, invoked by \citet{Micheli} to explain the non-gravitational acceleration $\boldsymbol{a}_{\rm ng}$ of 'Oumuamua, on its spin state. Torques due to outgassing can lead to changes of the direction \citep{Whipple} and magnitude \citep{Keller} of the spin. The latter has been measured for a number of Solar System comets as the variation of their spin period \citep{Mueller1996,Knight,Belton2011,Mottola,Bode}. Since both $\boldsymbol{a}_{\rm ng}$ and spin evolution are driven by the same underlying mechanism --- outgassing --- it is natural to expect the two outcomes to be related in some way.

\citet{RRR} has explored this connection by hypothesizing that the projection of the torque on the instantaneous spin axis of the comet $T_\Omega$ is related to the magnitude of the non-gravitational force $F_{\rm ng}$ via an "effective lever arm" $\zeta D$ as
\ba   
T_\Omega=\zeta D F_{\rm ng},
\label{eq:F}
\ea
where $D$ is the characteristic dimension of the object (e.g. radius for roughly spherical objects) and $\zeta$ is the dimensionless "lever arm" coefficient. This simple but intuitive prescription transforms the equation $I\dot\Omega=T_\Omega$ for the evolution of spin frequency $\Omega$ ($I$ being the moment of inertia with respect to spin axis) into \citep{RRR}
\ba  
\dot\Omega=\zeta\frac{D M}{I}a_{\rm ng},
\label{eq:spinev}
\ea
where $a_{\rm ng}=\boldsymbol{a}_{\rm ng}$ is the magnitude of the non-gravitational acceleration, $\boldsymbol{a}_{\rm ng}={\bf F}_{\rm ng}/M$. 

Introduction of the dimensionless lever arm parameter $\zeta$ effectively absorbs our ignorance of many aspects of the cometary outgassing: geometry of mass loss, shape of the object, etc. Some symmetric models of mass loss, e.g. those with the reactive force passing through the rotational axis of the object at every point on the surface \citep{Whipple,Sekanina1984}, naturally result in $\zeta=0$. On the other hand, if the reactive force is constant in magnitude and always normal to the surface, then ${\bf F}_{\rm ng}$ is identically zero, while $T_\Omega$ does not have to vanish if the comet has irregular shape, meaning $\zeta\to\infty$. 

\citet{RRR} used equation (\ref{eq:spinev}) to determine the values of $\zeta$ for 7 comets, which have measurements of both spin rate change per orbit $\Delta\Omega$ and non-gravitational acceleration $a_{\rm ng}$. He found a strong linear correlation between $\Delta\Omega$ and $a_{\rm ng}$ (the radial dependence of $a_{\rm ng}$ was modeled using the conventional prescription of \citet{Marsden}). The basic characteristics of the distribution of $\log\zeta$ for this sample (mean and dispersion) are \citep{RRR}
\ba  
\langle\log\zeta\rangle=-2.21,~~~\sigma_{\log\zeta}=0.54,
\label{eq:stats}
\ea  
implying rather small spread. Values of $\zeta$ corresponding to $\langle\log\zeta\rangle$, and $\langle\log\zeta\rangle\pm\sigma_{\log\zeta}$ are quite low: $\zeta=0.006$, $0.0017$, and $0.021$, respectively. Note that, because $\Delta\Omega$ is an integral characteristic (change of $\Omega$ accumulated over the full orbit), the values of $\zeta$ determined in this way effectively represent {\it averages} of instantaneous $\zeta$ defined by equation (\ref{eq:F}) over an orbit.


\subsection{Spin evolution of 'Oumuamua}  
\label{sect:spin_Ou}


Encouraged by the results of \citet{RRR}, we now apply this framework for assessing spin period variability of 'Oumuamua. More specifically, we use the measurement of its non-gravitational acceleration to estimate the spin rate change $\Delta\Omega$ accumulated along its trajectory through the Solar System. While this exercise is very similar to what has been done in \citet{RRR}, there are a couple of important differences.

First, unlike Solar System comets, 'Oumuamua moves on a hyperbolic trajectory. To characterize its motion through the Solar System we adopt the velocity at infinity $v_\infty=26~\mbox{km s}^{-1}$ and periastron distance $r_p= 0.26$ au. Conservation of energy and angular momentum leads to the following expression for the radial velocity $v_r$  at a distance $r$ from the Sun:
\ba   
v_r(r)=v_\infty r^{-1}\sqrt{(r-r_p)\left[r+r_p(1+\Theta)\right]},
\label{eq:hyperbolic}
\ea  
where $\Theta=2GM_\odot/(r_pv_\infty^2)\approx 10.2$. At 1 au the radial velocity of 'Oumuamua is $v_r($1 au$)=44$ km s$^{-1}$.

Second, non-gravitational acceleration of comets is usually modeled using the radial profile of $a_{\rm ng}$ suggested by \citet{Marsden}, which is motivated by the expected rate of sublimation of water ice. At the same time, fit to the astrometric data attempted by \citet{Micheli} used a non-standard expression for the function $g(r)$, different from that of \citet{Marsden}, see equation (\ref{eq:accel}).

Keeping this in mind, we now go back to equation (\ref{eq:spinev}). It is natural to associate characteristic dimension $D$ of 'Oumuamua with its long semi-axis $a$. Substituting the expression (\ref{eq:accel}) into equation (\ref{eq:spinev}) and using moment of inertia $I$ around its minor axis (see \S \ref{sect:rot}) one finds
\ba  
\dot\Omega=5\zeta\frac{A_1 a}{a^2+b^2}r_1^{-2}.
\label{eq:spinevOu}
\ea
Integrating this equation over $dt=dr/v_r$ from $r=r_p$ to $r=\infty$ using the expression (\ref{eq:hyperbolic}) and doubling the result (to account for both the incoming and outgoing parts of the trajectory) one finds the full change of the spin rate of 'Oumuamua during its passage through the Solar System to be 
\ba
\Delta\Omega=\zeta\Delta\Omega_1.
\ea  
Here 
\ba   
\Delta\Omega_1 &=& 
10\frac{(\mbox{au})^2A_1 a}{(a^2+b^2)v_\infty r_p} \int\limits_1^\infty \frac{x^{-1}dx}{\sqrt{(x-1)\left(x+1+\Theta\right)}}\nonumber\\
&\approx & 3.62~\mbox{s}^{-1}.
\label{eq:dO-Ou}
\ea   
is the value of the spin rate change if the lever arm parameter $\zeta$ were equal to unity. 

This calculation assumes that the $a_{\rm ng}(r)$ profile given by equation (\ref{eq:accel}) holds along the full orbit of 'Oumuamua through the Solar System. However, this fit was established by \citet{Micheli} using astrometric data acquired only between 1 au and 3 au. Thus, it makes sense to also evaluate the change of $\Omega$ accumulated over just this outgoing portion of the orbit. Proceeding analogous to the calculation of $\Delta\Omega$ (but integrating only between 1 and 3 au) one finds 
\ba   
\Delta\Omega^{\mbox{(1-3)au}}=\zeta\Delta\Omega_1^{\mbox{(1-3)au}},~~~
\Delta\Omega_1^{\mbox{(1-3)au}}=0.27 ~\mbox{s}^{-1}.
\ea   

Let us now assume for simplicity that the value of the 'Oumuamua's dimensionless lever arm $\zeta$ is comparable to that found for the Solar System comets (this is a very conservative assumption given the object's highly asymmetric shape, see \S \ref{sect:disc_Ou}). Then, adopting $\zeta=10^{\langle\log\zeta\rangle}$, see equation (\ref{eq:stats}), one finds 
\ba
\Delta\Omega=0.022~\mbox{s}^{-1},~~~\Delta\Omega^{\mbox{(1-3)au}}=0.0016~\mbox{s}^{-1}.
\label{eq:values}
\ea
These values correspond to spinning up an initially non-rotating object to very short periods of $\approx 5$ min and $\approx 1$ hr, respectively.


\section{Implications for 'Oumuamua}  
\label{sect:disc_Ou}

\begin{figure}
\centering
\includegraphics[width=0.5\textwidth]{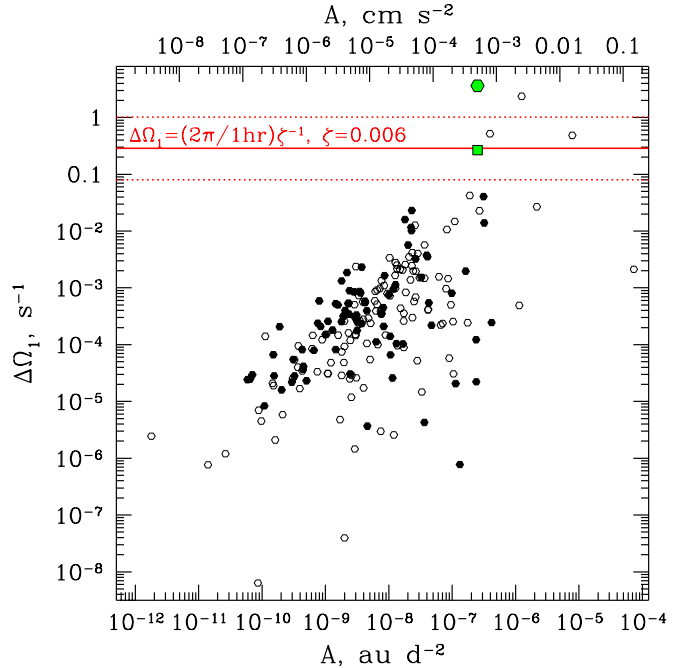}
\caption{
Change of the spin rate due to non-gravitational forces accumulated over a single orbit $\Delta\Omega_1$ (assuming lever arm parameter $\zeta=1$), computed for a sample of 209 Solar System comets with available data on the amplitude of the non-gravitational acceleration. This parameter measures object's potential for rapid spin evolution; $A$ is the normalization of $\boldsymbol{a}_{\rm ng}$ at 1 au, assuming $a_{\rm ng}(r)$ profile from \citet{Marsden}. Filled hexagons represent objects with measured sizes; empty ones have no size information available and we set radius $R=10$ km for them. Green hexagon and square represent 'Oumuamua (we set $A=A_1$ for it) and correspond to $\Delta\Omega_1$ and $\Delta\Omega_1^{\mbox{(1-3)au}}$. Horizontal solid curve describes rotation rate change of $2\pi$ hr$^{-1}$ (i.e. spin-up to 1 hr period from a non-rotating state) for $\zeta=0.006$ (corresponding to the mean of $\log\zeta$); dashed lines illustrate $1\sigma$ deviation in $\log\zeta$, see equation (\ref{eq:stats}). See text for more details.
\label{fig:comets}}
\end{figure}

It is clear that both $\Delta\Omega$ and $\Delta\Omega^{\mbox{(1-3)au}}$ are very large. To better put these values into context, we compare our expectations for spin evolution of 'Oumuamua with a large sample of Solar System comets previously studied by \citet{RRR}. Figure \ref{fig:comets} shows $\Delta\Omega_1$ computed for a sample of 209 Solar System comets with measured non-gravitational accelerations. The data on $A$ --- amplitude\footnote{Calculation of $\Delta\Omega_1$ for these comets uses $g(r)$ profile from \citet{Marsden}.} of $\boldsymbol{a}_{\rm ng}$ at 1 au, orbital parameters and sizes (when available, filled symbols; for objects with no size information we assign a radius of 10 km, open symbols) of comets come from the JPL Small Body Database\footnote{https://ssd.jpl.nasa.gov/?comets}. Green dots at the very top of the plot represent $\Delta\Omega_1$ and $\Delta\Omega_1^{\mbox{(1-3)au}}$, correspondingly, see equation (\ref{eq:values}). This figure is similar to Figure 3 of \citet{RRR}.


One can see that 'Oumuamua's spin rate change per orbit $\Delta\Omega_1$ (green hexagon) exceeds that of any Solar System object. This makes this interstellar object unique in yet another category. Its value of $A=A_1$ --- non-gravitational acceleration at 1 au --- is also among the highest in our sample \citep{Micheli}. Even considering the change of $\Omega$ between 1 and 3 au (i.e. $\Delta\Omega_1^{\mbox{(1-3)au}}$, green square), 'Oumuamua still ends up squarely among a handful of objects most susceptible to spin variations --- only three comets (discussed in detail in \citealt{RRR}) have $\Delta\Omega_1$ larger than $\Delta\Omega_1^{\mbox{(1-3)au}}$. 

To put these comparisons into absolute terms, we assume for now that 'Oumuamua's outgassing asymmetry is not too different from that of the Solar System comets and set $\zeta=0.006$, corresponding to the mean of $\log\zeta$ found in \citet{RRR}, see equation (\ref{eq:stats}). This allows us to relate $\Delta\Omega_1$ in Figure \ref{fig:comets} to the actual spin rate variation $\Delta\Omega=\zeta\Delta\Omega_1$. 

An important value of $\Delta\Omega$ is the one that would result in rotational fission of an object. Using results of \citet{Davidsson} we find that, in the absence of internal strength (i.e. if the object is held by gravity alone), rotational breakup of 'Oumuamua would happen at longer critical period $P_{\rm crit}\approx 9.8$ hr (for bulk density of 1 g cm$^{-3}$) than for a spherical object with the same density and mass, for which $P_{\rm crit}\approx 3.3$ hr (see also \citealt{Meech} and \citealt{McNeill}). This is a consequence of the highly elongated shape of 'Oumuamua. However, if we account for non-zero tensile strength of the object according to the prescription of \citet{Davidsson}, we find $P_{\rm crit}< 1$ hr. 

Unfortunately, neither density nor tensile strength are known for 'Oumuamua. For this reason we opted to assume, rather conservatively, that breakup happens at spin period of $P_{\rm crit}=1$ hr (although the critical period is likely longer). An object starting as slowly spinning would need to acquire $\Delta\Omega=2\pi P_{\rm crit}^{-1}=2\pi$ hr$^{-1}$ to get to $P_{\rm crit}$. Corresponding value of $\Omega_1$, assuming $\zeta=0.006$, is shown as a horizontal solid line in Figure \ref{fig:comets}. Dotted lines correspond to values of $\zeta$ deviating from $10^{\langle\log\zeta\rangle}$ by $\sigma_{\log\zeta}$, see equation (\ref{eq:stats}). 

One can see that 'Oumuamua's $\Delta\Omega_1$ exceeds the limit of rotational fission (even when the low value of $\zeta=0.0017$ is adopted). Even spin-up between only 1 and 3 au would still get the rotation period down to $P_{\rm crit}$ (green square falls on top of horizontal solid line). Thus, outgassing-driven spin evolution of 'Oumuamua is expected to result in its rotational fission at some point along its orbit, as has been observed for some Solar System objects \citep{Jewitt2016,Jewitt2017}. 

Another useful way to characterize spin variability of 'Oumuamua's is via its instantaneous rotational evolution timescale $\tau_\Omega=|\Omega/\dot\Omega|$. Using equation (\ref{eq:spinevOu}) we can write $\tau_\Omega$ as
\ba   
\tau_\Omega=
\frac{2\pi}{5\zeta}\frac{a^2+b^2}{P a A_1}r_1^2\approx 4~\frac{0.006}{\zeta}r_1^2~\mbox{d},
\label{eq:spin1}
\ea   
where $P=2\pi/\Omega=8.67$ hr is the adopted rotation period of 'Oumuamua. 
 
Alternatively, by looking in some detail at the actual physics of outgassing \cite{Jewitt1997} suggested the following expression for the spin evolution timescale:
\ba  
\tau_\Omega\approx \frac{I\Omega}{k_T D\dot M v_{\rm th}},
\label{eq:spin2}
\ea
where $\dot M$ is the mass loss rate due to outgassing and $v_{\rm th}$ is the speed of escaping gas (as before $I$ and $D$ are the moment of inertia and characteristic size). Dimensionless asymmetry parameter $k_T$ is closely related to our lever arm parameter $\zeta$; $k_T=1$ and 0 for purely tangential and central mass loss, respectively. 

\citet{Jewitt1997} advocated using $k_T=0.05$. Adopting this value, as well as $v_{\rm th}=0.5$ km s$^{-1}$ and $\dot M=10^4$ g s$^{-1}$, used by \citet{Micheli} to explain the high value of $a_{\rm ng}$ for 'Oumuamua, we find $\tau_\Omega\approx 14$ hr. This is almost an order of magnitude lower than the estimate (\ref{eq:spin1}). This discrepancy suggests that $k_T=0.05$ adopted by \citet{Jewitt} is a significant overestimate of the asymmetry of outgassing, something that has been noted already by \citet{RRR}. A different estimate $0.005<k_T<0.04$ obtained by \citet{Belton2011} based on observations of 9P/Tempel 1 leads to better quantitative agreement between the estimates (\ref{eq:spin1}) and (\ref{eq:spin2}).

Regardless of the method used to derive $\tau_\Omega$, its short value again implies that spin period of 'Oumuamua should be changing very rapidly. Based on (\ref{eq:spin1}), for $\zeta=0.006$ one expects a change of $P$ of more than an hour to occur in less than a day! Within several weeks torques due to outgassing should have spun 'Oumuamua up to breakup. However, this has, obviously, not happened. 

Moreover, despite the complications related to the tumbling motion of 'Oumuamua, most recent analysis \citep{Belton} of its photometric variability using observations extended over a period of about a month reports $P=8.67$ hr, close to many previous determinations \citep{Bolin,Bannister,Drahus,Fraser}. If we interpret the formal uncertainty of this measurement ($\pm 0.34$ hr) as resulting from outgassing-driven spin variability, then this would imply $\tau_\Omega\gtrsim 2$ yr, much longer than suggested by our estimate (\ref{eq:spin1}).  

A naive way to resolve this discrepancy is to suppose that $\zeta=0.006$ adopted in our calculations \citep{RRR} is an overestimate of the 'Oumuamua's lever arm parameter. However, according to equation (\ref{eq:spin1}), to get $\tau_\Omega$ up to 2 yr would require $\zeta\approx 3\times 10^{-5}$, an extremely small value. The physical size of the corresponding lever arm would then be $\zeta D\approx 0.7$ cm for the object's size $D=a=230$ m. Such degree of symmetry, resulting in almost complete cancelation of the torques while leaving a substantial linear acceleration $a_{\rm ng}$, is highly improbable. In fact, based on the lack of Solar System comets with very high values of $\Delta\Omega_1\gtrsim 1$ s$^{-1}$ (see Figure \ref{fig:comets}) \citet{RRR} has argued for the existence of a lower limit on $\zeta$ of about $10^{-3}$, still much higher than the value quoted above.

Moreover, given the extreme elongation of the 'Oumuamua compared to the vast majority of the Solar System objects, one should, in fact, expect its $\zeta$ to be substantially {\it higher} than $10^{\langle\log\zeta\rangle}=0.006$ typical for the Solar System comets. This would {\it reduce} $\tau_\Omega$ given by equation (\ref{eq:spin1}), further increasing tension with the observed lack of spin evolution for this object. 

\citet{Belton} raised a possibility that 'Oumuamua may be a highly oblate spheroid in a long-axis mode of rotation. This would not change our spinup timescale estimate (\ref{eq:spin1}) and other conclusions appreciably as the moment of inertia for this rotational state is essentially the same as for a cigar-shaped object spinning around its short axis. One could also argue that 'Oumuamua's tumbling motion represented by precession around the long axis (if it is cigar-shaped) with a period of $\approx 54.5$ hr \citep{Belton} would result in a periodic torque flip, slowing down (or at least complicating) object's evolution towards rotational breakup. However, the period of such precession is comparable to $\tau_\Omega$ given by the equation (\ref{eq:spin1}) so that variations of 'Oumuamua's spin period should have been observable even during a single precessional cycle. Also, period of precession around long axis would be changing rapidly as well (given the low moment of inertia around this axis), likely resulting in a very chaotic evolution of the spin, which is not supported by observations \citep{Belton}. 

Our conclusions are also relatively robust with respct to the uncertainties in the determination of various parameters of 'Oumuamua. For example, they are insensitive to its bulk density, which is essentially unknown. Also, the dimensions of the object scale with its (unknown) albedo $p$ as $\propto p^{-1/2}$. Thus, a change of an adopted value of $p$ by a factor of 4 would result in variation of our estimates (\ref{eq:dO-Ou}) and (\ref{eq:spin1}) by only a factor of 2. This would have little effect on our conclusions. 

Calculations presented in this work make it obvious that the large magnitude of the observed anomalous acceleration of 'Oumuamua is difficult to explain via its cometary activity, i.e. by intense outgassing from its surface. Our arguments against cometary activity as the driver of the anomalous acceleration of 'Oumuamua are quite fundamental and are based on the conservation of angular momentum\footnote{Similar arguments are independently advanced by Guzik \& Drahus in their abstract (301.05) for the upcoming DPS meeting.} for this object. The high intensity of outgassing implied by the measurement of $a_{\rm ng}$ by \citet{Micheli} is also in drastic contrast with the lack of clear signs of mass loss from 'Oumuamua, despite multiple targeted observations aimed at addressing this issue \citep{Knight17,Jewitt,Ye,Meech}. As a result, we believe that the classification of 'Oumuamua as an interstellar asteroid (rather than a comet) is more robust and justified by observations at the moment. The puzzle of its anomalous acceleration would need to be resolved without invoking powerful outgassing from its surface (we do not suggest an alternative mechanism in this work).

This conclusion would have significant implications for understanding the origin of this unique object. Indeed, refractory asteroids are far more difficult to detect than the volatile-rich, cometary objects \citep{Engel}. The asteroidal nature of 'Oumuamua would imply that refractory asteroids dominate the Galactic population of free-roaming $0.1-1$ km objects, in agreement with some theories for the origin of 'Oumuamua \citep{Cuk,Rafikov}.


\section{Summary}  
\label{sect:summary}


In this work we explored the implications of a recent measurement of non-gravitational  acceleration $a_{\rm ng}$ of the interstellar minor object 'Oumuamua \citep{Micheli} for its rotational state. Our main conclusions are as follows. 

\begin{enumerate}

\item Torques caused by outgassing (the same process that gives rise to the claimed $a_{\rm ng}$) should drive rapid spin evolution of 'Oumuamua on a characteristic timescale of several days. This conclusion assumes that reactive torques and $a_{\rm ng}$ are related via the dimensionless lever arm parameter typical for the Solar System comets, $\zeta=0.006$ \citep{RRR}.

\item Such fast spin evolution is incompatible with the observed stability of the object's spin rate on month-long timescales. Matching observations would require extreme degree of symmetry of outgassing (leading to torque cancellation), $\zeta\lesssim 3\times 10^{-5}$, which is implausible.

\item Moreover, given the amplitude of its claimed non-gravitational acceleration, 'Oumuamua should have been spun up to the limit of rotational fission and fragmented during its travel through the Solar System.

\item That this did not happen, as well as the observed spin stability of 'Oumuamua, strongly argue against its cometary activity (i.e. powerful outgassing from its surface), which was invoked to explain its claimed $a_{\rm ng}$ in the first place \citep{Micheli}. 
    
\end{enumerate}

Based on these arguments, as well as the lack of direct indicators of outgassing, we conclude that the classification of 'Oumuamua as a volatile-rich comet is dubious and raises more questions than it provides explanations. We encourage further analyses of the astrometric data to better characterize its anomalous acceleration \citep{Micheli}. Settling the issue of whether this unique object has asteroidal or cometary composition is crucial for understanding its origin \citep{Rafikov}. 

\acknowledgements

I thank an anonymous referee for prompt and constructive review and for bringing the DPS abstract of Guzik \& Drahus to my attention. Useful comments by Scott Tremaine, David Jewitt and Matthew Knight are gratefully acknowledged. Financial support for this study has been provided by NSF via grant AST-1409524 and NASA via grant 15-XRP15-2-0139.



\bibliographystyle{apj}
\bibliography{references}

\end{document}